# INTELLIGENT DEVICE USED BY AN INFORMATION SYSTEM FOR IDENTIFYING AND MONITORING OF PATIENTS

Ioan Ungureanu[1], Cristina Elena Turcu[2], Cornel Turcu[3], Vasile Gheorghita Gaitan[4]

[1-4]Ştefan cel Mare University of Suceava

e-mail: {ioanu32@yahoo.com, cristina@eed.usv.ro, cturcu@eed.usv.ro, gaitan@eed.usv.ro}

**Abstract** — The development of medical services quality by means of integrating certain software applications, allowing the decrease of errors within the diagnosis and establishment of patients' treatments, has been taken into considerations for the time being. The aim of this paper consists in defining the hardware and software architecture of an embedded system, based on RFID technology, in order to identify patients and to achieve real time information concerning the patients' biometric data, which might be used in different points of the health system (laboratory, family physician, etc.).

**Keywords:** *embedded system, real time, microprocessor, tag, RFID reader, CIP card, web server.*

## 1. INTRODUCTION

The level of applying informatics within the healthcare system is relatively reduced for the time being in Romania, where information about patients has not been shared at the level of medical entities, where the medical records of patients are not unitary and complete, and cannot be accessed online by the medical staff, when in need. In this context our research team proposed an integrated system for identification and monitoring of patients – SIMOPAC [1]. This system was design to integrate within the distributed medical information system, and privately, to solve the problems related to identification and monitoring the patients. The SIMOPAC system will assure the information exchange with electronic health record (EHR/EMR) [2][3] systems set up to healthcare units in accordance with the HL7 [4] standards specifications. Within the SIMOPAC system, the access to medical services is suggested to be carried out by means of information stored within a *Personal health Information Card* (CIP) [1]. This card will be implemented by using the RFID technologies [5], where information carrier is represented by a transponder (tag) – an electronic memory-based chip.

An embedded device with an RFID reader can be used in order to identify a patient (Figure 1). This device will read the patient's CIP and will carry out its interpretation. The contents of the patient CIP card [1] can be seen by means of a display, which is endowed to the mobile device or by means of the local Ethernet network (a mini WEB server will operate on this device). If supplementary information is asked, the device will be connected to the Internet and the application of the server indicated by URI (Uniform Resource Identifier) [1] on CIP will be started. The device will transmit to the SIMOPAC system the IP and the serial number of the CIP, for which the information was asked. The advantage proved by this device is represented by its mobility. This device can also read information from different medical devices (for instance, the blood pressure monitor), corresponding to a patient, being able to transmit it towards the SIMOPAC system of the healthcare unit level. The development of a low cost device is aimed to be achieved. The production costs for such device have to be below 600 Euro (the cost of a PDA with RFID reader and wireless network). In this paper, two possible solutions are presented, one allowing the accomplishment of complex operations and one being able to carry out just the file inspection of the CIP card.

## 2. THE RFID TECHNOLOGY

Radio Frequency Identification (RFID) [6][7] uses the wireless technology, in order to identify entities – such as the products belonging to a supply chain, wild or domestic animals, which are about to be tracked, persons, etc. – from a distance, where a direct visibility is not possible [7]. Thus, within a supplying chain, RFID can be considered as a wireless solution towards the bar codes optical scanning, but with certain essential differences. The RFID technology can be used in healthcare for patient care (patient identification and tracking, caregiver verification, medication verification) and device tracking. The RFID tags include a series of desired information, related to the specific applications. The tag, either active, passive or semi-active, can be automatically read/write by an RFID reader, which is able to transmit the tag information to a host device (computer or embedded device), in order to be processed and/or stored.

The systems based upon the RFID technology are generally composed of three components:
- an RFID reader;
- an RFID tag;
- a computer or any other data processing system.

The potential benefits of the RFID technology [6][7] proved to be multiple. Thus, for example, in supply chain RFID-enabled systems help companies improve customer service, cut costs, reduce labour, increase accuracy, and improve production throughput. In healthcare industry the RFID technology is used for tracking movements in hospitals (doctors, nurses, patients, visitors or any movable valuable equipment). In the U.S., for example, some health-care providers are already using RFID-based systems to track equipment, instruments and sponges used in surgery to ensure that nothing is left behind inside a patient. And some health-care providers are already using RFID-enabled labels to track specimens and laboratory results to ensure that they are not misplaced, as well as tracing pharmaceutical products to ensure that the correct medication and dosage is being administered.

More than 400 U.S. hospitals are currently using RFID-based, baby-and-mother matching systems to prevent mix-ups and abductions – the RFID system triggers a lock-down if an infant is removed from a secured area without authorization [8][9]. Also, it is believed to be the best tool to fight drug counterfeiters.

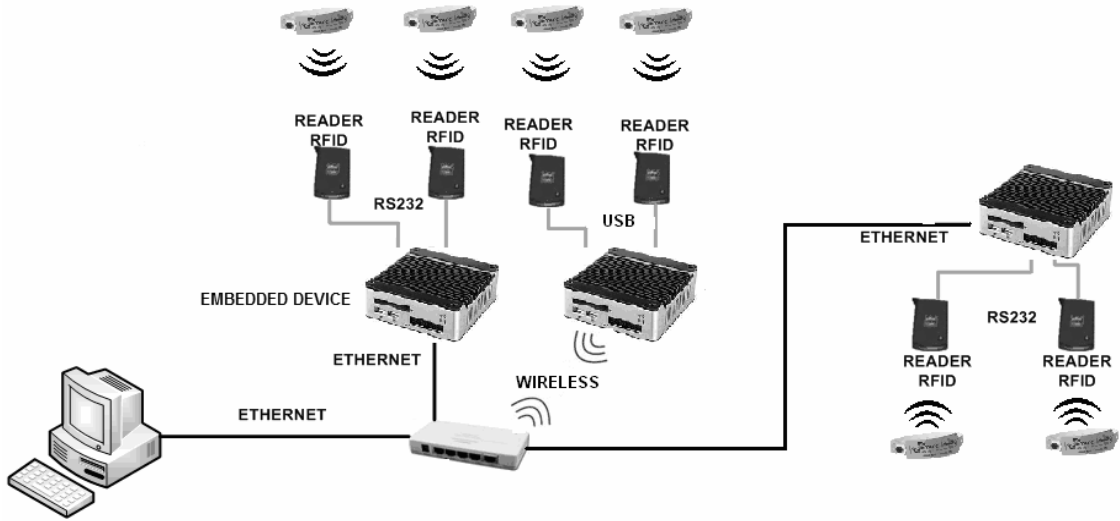

**Figure 1. Connecting the intelligent device to a network.**

## 3. MULTI-AGENT TECHNOLOGY

Agent technology is an emerging and promising research area, which increasingly contributes to the development of value-added information systems for different applications. An agent is a small, autonomous or semi-autonomous software program that performs a set of specialized functions to meet a specific set of goals, and then provides its results to a customer (e.g., human end-user, another program) in a format readily acceptable by that customer [10]. For example, agent technology has been applied in the area of extracting information from heterogenous data sources in the World Wide Web. The performance evaluation of the agent-based system versus traditional systems (client-server and relational database based systems) was undertaken by some researchers [11][12]. The tests reveal that the agent approach provides better response times as well quicker notification processing.

Healthcare systems are characterized by wide variety of applications that work in separated and isolated environments. The use of agent technology in healthcare system has been on the increase in the last decade. Thus, multi-agent systems do have an increasingly important role to play in health care domains, because they significantly enhance our ability to model, design and build complex, distributed health care software systems [13].

## 4. THE AIM OF THE INTELLIGENT DEVICE

The aim of this paper is represented by defining the hardware and software architecture of an intelligent embedded subsystem, in order to achieve real time information concerning the patients' biometric data, which might be used in different points of the healthcare system (laboratory, family physician, etc.)

Before defining the potential solutions, the functional and non-functional requirements of the hardware subsystem have to be defined. Thus, it has to carry out the following functions:
- to allow the connection of an RFID reader, by using an USB port or a serial port (RS232 or RS485);
- to allow the sending/quering of data towards the central database;
- to allow the visualisation of patients' data (the CIP content), by means of a monitor/display;
- to allow reading the information from the medical devices;
- to send alarm;
- to allow the communication with the SIMOPAC system, in order to achieve supplementary information about a patient;
- to modify the CIP card content;
- to assure a high-level of security.

## 5. PROPOSED SOLUTIONS

The first proposed solution refers to the embedded computer eBox2300SX [14], that connects RFID readers, medical devices and various sensors which can detect the patients' vital signs, such as electrocardiogram (ECG), and body temperature. This solution might be developed by using the development kit eBox2300SX, presented in Figure 2. This is a computing embedded system, of low cost, based upon the x86 processor.

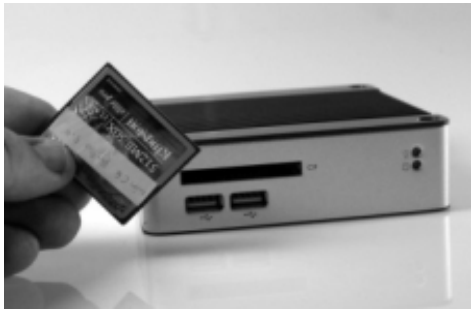

**Figure 2. The eBox 2300SX kit.**

This includes enough specific features towards a desktop PC. The Vortex86 processor is compatible to X86 family, being of SoC (System on Chip) type. The power consumption is reduced (about 15W towards those 700W for the desktop PC), but the computing ability is less than that of a PC. It has 128MB DRAM, an AMI type PC BIOS, and an internal flash memory of 256MB. Within this configuration, operating systems as Linux embedded and Windows CE can operate. Operating systems as Windows XP and Windows Vista can be set up also, where the system might be able to operate as desktop type system, but having limited resources.

The *advantages* of the solution:
- high storing capacity – an HDD can be set upon;
- execution of real time tasks by using Windows CE operating system;
- capability to operate as "mini web server", which can be used in order to remotely visualise the patients records;
- using of agents to transmit towards the physician the analysis results and alarms information;
- local visualisation of recordings corresponding to a patient – the embedded device can be used with a monitor connected to VGA output;
- capability of connection with RFID readers at 3 USB ports and 2 serial ports;
- low price.

The *disadvantage* of the solution:
- it is not possible to use a display; only a monitor can be attached.

The use case diagram for the application that will be carried out on eBox 2300 is presented in Figure 3. On this system, Windows CE 6 operating system will be set up.

The application will use the information from a tag (existing on a patient's hand), through an RFID reader. Upon basis of information read, immediate medical data can be achieved (blood group, RH, allergic substances, HIV/AIDS, or other chronic or transmittable diseases, etc.).

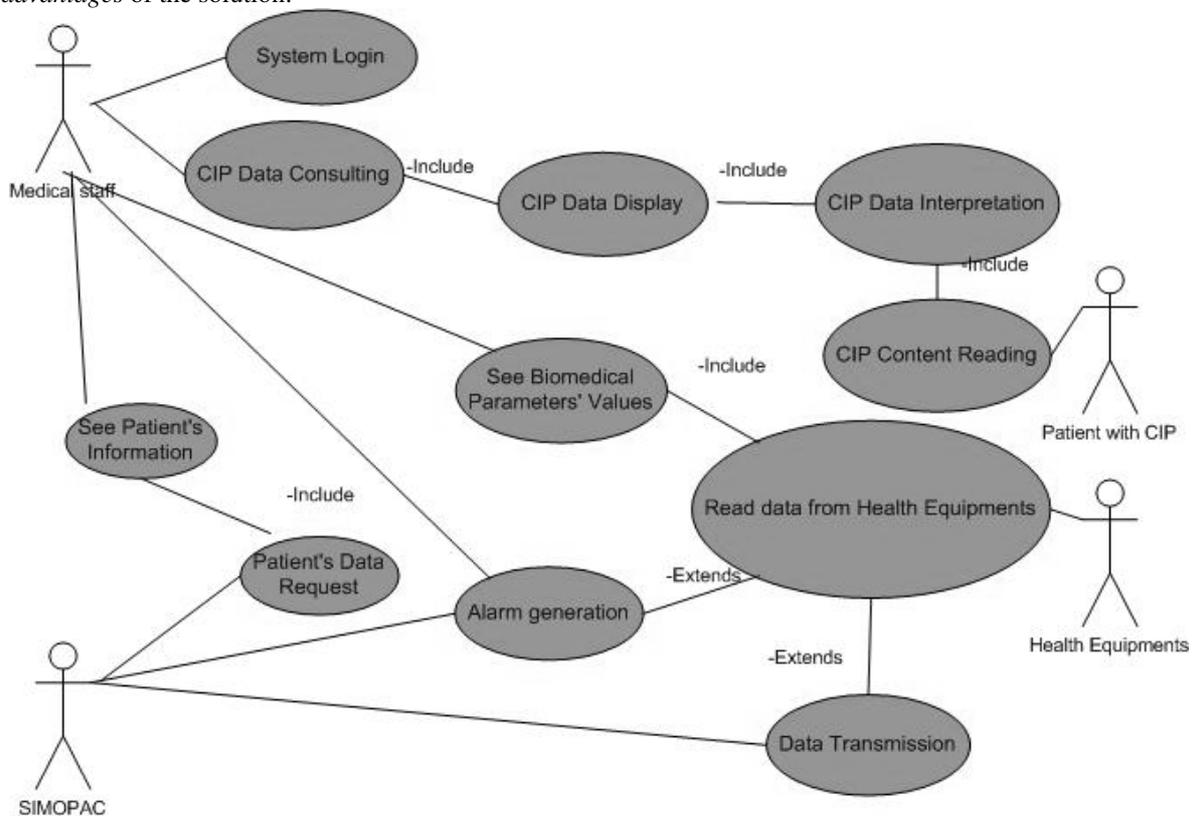

**Figure 3. The use case diagram of the intelligent device.**

In order to achieve supplementary information, the application will be connected through the Internet to the server indicated by URI on CIP (on the tag). The user will indicate the preferred language, of certain potential on that server. The SIMOPAC server processes only the authorised request for accessing an electronic medical record of a patient and returns the answer. Within this application, the content of CIP card can be modified (any modification can be accomplished only after the user will log on). Information can also be read by means of this application, from the medical devices that have a Bluetooth communication interface (by using an USB – Bluetooth interface). This information can be saved into the CIP card or send to the SIMOPAC system. This is because the eBOX 2300 system can function in two ways: with or without monitor, and the application operates in two ways: as graphical application for the monitor solution and as mini web server, in order to be accessed by means of local Ethernet network.

We propose an agent-based approach that will provide a high degree of flexibility and security. In our system, each human actor (physician, patient) will be assigned a specific agent. We will define the following agent categories:
- physician agent;
- biometric agent;
- patient agent;
- SIMOPAC agent.

We briefly present specific task performed by agent categories:
- Physician agent
   o notify the physician that biometric results are available;
   o receive test results from the biometric agent;
   o receive alarm messages;
   o display test results data to the physician;
   o send test result to SIMOPAC agent;
   o receive and display desired information from SMOPAC agent.
- Biometric agent
   o read the connected medical devices and sensors which can detect the patients' vital signs, such as electrocardiogram (ECG), and body temperature;
   o process the medical information;
   o notify the physician agent that results are available;
   o send alarm messages as soon as the abnormal results are detected;
- Patient agent
   o identify the patient;
   o read the CIP content;
- SIMOPAC agent
   o construct and send HL7 messages;
   o send the desired information to physician agent;
   o update the SIMOPAC database with test results.

The considered agents will enable integration of our device in different healthcare systems by providing transparent and secure communication mechanisms based on standard protocols utilized within the healthcare area. Another potential solution might be carried out by Rabbit processor [15], to which RFID readers and medical devices can be connected (Figure 5). This solution can be accomplished in two ways:
- with wireless connection, if RCM4400W [15] is used;
- with Ethernet connection, if RCM3000 [15] is used.

A display having a keyboard (as seen in figure 6) can be attached to these modules.

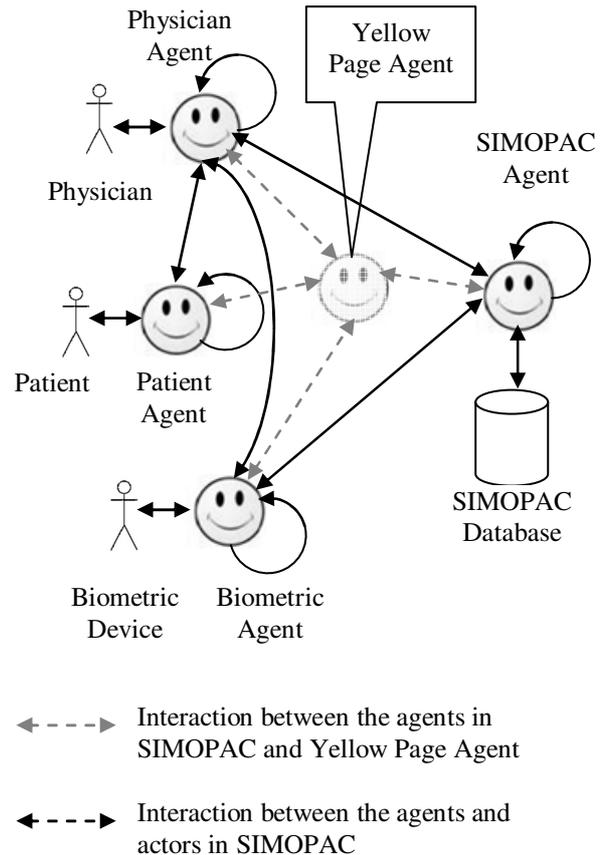

◀- - -▶ Interaction between the agents in SIMOPAC and Yellow Page Agent

◀- - -▶ Interaction between the agents and actors in SIMOPAC

**Figure 4. Agent overview.**

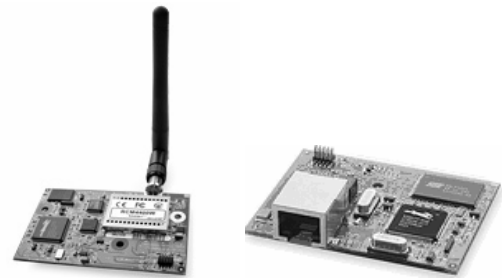

**Figure 5. RCM4400W RabbitCore and RCM3000 RabbitCore modules.**

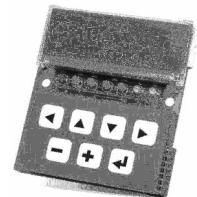

**Figure 6. Graphical LCD display with 7 keys.**

In order to determine the cost of solution, the prices related to motherboard, the power supply or RFID reader should be taken into consideration

The *benefits* of the solution:
- can use the µC/OS-II operating system – can use real time tasks;
- can operate as a "mini web server", which might be used for remotely visualisation of patients' records;

- allow the connecting of RFID readers at 6 serial ports;
- displaying the information, by using an LCD display and 7 keys;
- low price.

The *disadvantages* of the solution:
- low storing capacity;
- cannot execute complex tasks;
- limited memory of code and data.

In order to develop the software application on the device having RCM3000 module, the C language and Dynamic C compiler can be used. Dynamic C is a compiler especially created for the processors that are based upon Z80 processor (as Rabbit 3000 processor situation). This offers facilities of debugging towards source code or machine code level. It also includes libraries that allow the implementation of web servers, ftp servers, communication by means of sockets, communication by means of serial ports, using the files system for the flash memory, sending of e-mails. Dynamic C takes care of memory managing, but also allowing that memory to be managed by the programmers. In order to allow the real time tasks execution, the μC/OS-II operating system will be used.

## 6. CONCLUSIONS

As the RFID technology is becoming cheaper and cheaper, it is for sure that it would be used in everyday life soon. Our research team proposed to integrate the RFID technology in healthcare domain, by defining CIP card. Within this paper, the general architecture of an embedded system, used for file inspection of a CIP card read from the patient tags was defined. The embedded system is used in order to achieve real time information concerning the patients' biometric data, which might be used in different points of the healthcare system. This system proves a low cost and allows the modification of a CIP card by the physician. Two solutions, which might be used for developing an embedded system, were taken into consideration: the first related to eBOX2300SX system, based upon Vortex86SX processor, and the second related to RCM3000 system, based upon Rabbit 3000 processor.

## ACKNOWLEDGMENT

The research presented in this paper was supported within the framework of the Romanian Ministry of Education and Research, "PNCDI II, Partnerships" program, under Grant named "SIMOPAC – Integrated System for the Identification and Monitoring of Patient" no. 11-011/2007.

## REFERENCES

[1] C. Turcu, Cr. Turcu, "Sistem informatic integrat pentru identificarea si monitorizarea pacientilor – SIMOPAC", vol. *Sisteme Distribuite*, Decembrie, 2008, Suceava, Romania (in Romanian).
[2] Smaltz, Detlev and Eta Berner. *The Executive's Guide to Electronic Health Records*, 2007, Health Administration Press, pp. 03.
[3] Hallvard Lærum, MD, Tom H. Karlsen, MD, Arild Faxvaag,, "Effects of Scanning and Eliminating Paper-based Medical Records on Hospital Physicians' Clinical Work Practice", 2006, *Journal of the American Medical Informatics*, pp. 588–595.
[4] *, "Health Level Seven", https://www.hl7.org.
[5] Jonathan Collins, "RFID Remedy for Medical Errors", *RFID Journal*, 2004, Available at http://www.rfidjournal.com/article/view/961/1/1.
[6] G. Borriello, "RFID: Tagging the world", *Communication ACM* (Guest Editorial to RFID Special Issue), vol. 48, no. 9, Sep. 2005, pp. 34–37.
[7] Liu, K. Yang, C. Zhang, and Z. Wang, "A transponder IC for wireless auto identification system," in *Proceedings of the 5th International Conference ASIC*, Oct. 2003, vol. 2, pp. 1114–1116.
[8] *, RFID technology can help save lives in health-care sector – and preserve privacy: Commissioner Cavoukian and HP, Available at http://h41131.www4.hp.com/ca/en/pr/rfid-technology-can-help-save-lives-in-health-care-sector.html
[9] Ann Cavoukian, Victor Garcia, *RFID and Privacy: Guidance for Health-Care Providers*, 2008, Available at http://www.ipc.on.ca
[10] Daniel H. Wagner Associates, Inc., Software Agents, http://www.wagner.com/technologies/softwareagents/softwareagents.html
[11] G. Yamamoto, H. Tai, Performance evaluation of an agent server capable of hosting large numbers of agents, *Fifth International Conference on Autonomous Agents*, Montreal, Canada, pp 363-369, May 28-June 1, 2001
[12] Y. El-Gamal, K. El-Gazzar, M. Saeb, A Comparative Performance Evaluation Model of Mobile Agent Versus Remote Method Invocation for Information Retrieval, *Proceedings of World Academy of Science*, Engineering and Technology, vol. 21, 2007, ISSN 1307-6884
[13] John Nealon, Antonio Moreno, Agent-Based Applications in Health Care, Applications of Software Agent Technology in the Health *Care Domain* , Whitestein Series in Software Agent Technologies, Birkhäuser Verlag (2003), pp. 3-18.
[14] *, "eBox-2300SX-LS User's Manual", Available at: http://www.compactpc.com.tw
[15] http://www.rabbit.com/